# Optical-Based Thickness Measurement of MoO₃ Nanosheets

**Sergio Puebla** [1], **Antonio Mariscal-Jiménez** [2], **Rosalía Serna Galán** [2], **Carmen Munuera** [1] and **Andres Castellanos-Gomez** [1,*]

[1] Instituto de Ciencia de Materiales de Madrid (ICMM-CSIC), E-28049 Madrid, Spain; sergio.puebla@csic.es (S.P.); cmunuera@icmm.csic.es (C.M.)

[2] Laser Processing Group, Instituto de Óptica (IO, CSIC), Serrano 121, 28006 Madrid, Spain; antonio.mariscal@csic.es (A.M.-J.); rosalia.serna@csic.es (R.S.G.)

[*] Correspondence: andres.castellanos@csic.es

**Abstract:** Considering that two-dimensional (2D) molybdenum trioxide has acquired more attention in the last few years, it is relevant to speed up thickness identification of this material. We provide two fast and non-destructive methods to evaluate the thickness of MoO₃ flakes on SiO₂/Si substrates. First, by means of quantitative analysis of the apparent color of the flakes in optical microscopy images, one can make a first approximation of the thickness with an uncertainty of ±3 nm. The second method is based on the fit of optical contrast spectra, acquired with micro-reflectance measurements, to a Fresnel law-based model that provides an accurate measurement of the flake thickness with ±2 nm of uncertainty.

## 1. Introduction

Since the isolation of graphene by mechanical exfoliation in 2004 [1], the catalog of different 2D materials with complementary properties keeps growing [2–9]. Among them, wide bandgap semiconductor materials have attracted a great deal of attention due to their potential in optoelectronic applications [10] requiring electrically conductive materials that are transparent to visible light. Molybdenum trioxide (MoO₃) in its $\alpha$-phase is a van der Waals material with a monolayer thickness of ~0.7 nm [11,12] and a direct bandgap of approximately 3 eV [13–15], suitable for such applications [16]. It has been used in thin films to enhance the injection of holes in organic light-emitting diodes as a buffer layer [17,18], in organic photovoltaics [19], perovskite solar cells [20] and silicon solar cells [21], furthermore it can be used in gas sensors [15,22,23]. Moreover, MoO₃ is interesting because of its photochromic, thermochromic, electrochromic effects [24–28] and catalytic properties in the partial oxidation of methanol to formaldehyde [29–33]. Sub-stoichiometric MoO₃ quantum dots have been synthesized as surface-enhanced Raman scattering substrates [34]. Furthermore, this material displays an in-plane anisotropy of the crystal structure in its layered phase ($\alpha$-MoO₃) [16,35–38], which can be exploited to fabricate novel optical and optoelectronic devices [39–41], and anisotropic phonon polariton propagation along the MoO₃ surface has been observed [42].

A rapid and non-destructive method to measure the thickness of MoO₃ would be highly desirable for the further development of this line of research. This is precisely the goal of this manuscript: to provide a guide to evaluate the thickness of MoO₃ nanosheets (in the 0–100 nm range) by optical microscopy-based methods. We propose two complementary approaches: first, a coarse thickness estimation based on the apparent interference color of the flakes and, second, a quantitative analysis of the reflection spectra using a Fresnel law-based model.

## 2. Materials and Methods

$MoO_3$ flakes were grown by a simple physical vapor transport method carried out at atmospheric conditions, developed in Reference [35]. A molybdenum foil was oxidized by heating it up on a hotplate at 540 °C, then a silicon wafer was placed on top. The molybdenum oxide sublimed and crystallized on the surface of the Si wafer, at a slightly lower temperature, forming $MoO_3$ flakes. The $MoO_3$ grown by this method was characterized by X-ray photoemission spectroscopy (XPS) finding that it was composed of a single-phase fully oxidized $MoO_3$ [35]. These $MoO_3$ flakes can be easily lifted from the Si wafer surface with a Gel-Film (WF x4 6.0 mil, from Gel-Pak) viscoelastic substrate and subsequently transferred to an arbitrary target substrate. We transferred the flakes onto silicon chips with a 297-nm $SiO_2$ capping layer (see the Supporting Information for the quantitative determination of the $SiO_2$ thickness) as it is one of the standard substrates for work with 2D materials.

### 2.1. Atomic Force Microscopy (AFM)

AFM characterization was performed at ambient conditions using two commercial AFM systems: (1) a Nanotec AFM system has been used [43] in dynamic mode with a NextTip (NT-SS-II) cantilever (resonance frequency of 75 kHz), (2) an ezAFM (by Nanomagnetics) AFM operated in dynamic mode with Tap190Al-G by Budget Sensors AFM cantilevers (force constant 48 $Nm^{-1}$ and resonance frequency 190 kHz).

### 2.2. Optical Microscopy and Spectroscopy

Optical microscopy images were acquired using a Motic BA310 Me-T microscope (Motic, Barcelona, Spain) (equipped with a 50× 0.55 NA objective and an AMScope MU1803 CMOS Camera) and reflection spectra were collected from a spot of ~1.5–2 μm diameter with a Thorlabs CCS200/M fiber-coupled spectrometer (Thorlabs Inc., Newton, New Jersey, United States). More details about the micro-reflectance setup can be found in Reference [44].

## 3. Results and Discussion

Figure 1a shows the atomic force microscopy (AFM) topography of eight $MoO_3$ flakes with different thicknesses, and Figure 1b shows their corresponding optical microscopy images. The direct comparison between the AFM and the optical images allows us to build up a color-chart correlating the apparent color of the $MoO_3$ flakes deposited on top of the 297 nm $SiO_2$/Si substrate (the $SiO_2$ thickness is experimentally measured by reflectometry with ±0.5 nm uncertainty) with their corresponding thickness with an uncertainty of ±3 nm. Similar approaches have been reported for graphene [45], transition metal dichalcogenides [46], $TiS_3$ [47] and franckeite [48]. This method has the main limitation that it requires the use of a specific $SiO_2$ thickness as the interference colors of the $MoO_3$ flakes strongly depend on the underlying substrate. In this work, we provide color-charts for four different nominal $SiO_2$ capping layer thicknesses: 297 nm, 271 nm, 148 nm and 88 nm. Figure 2 shows the correlation of optical images and AFM images for $MoO_3$ flakes transferred onto an 88 nm $SiO_2$/Si substrate. We address the reader to the Supporting Information for the data corresponding to the 148 nm and 271 nm thick $SiO_2$ substrates, named as Figures S1 and S2, respectively . Figure 3 shows a comparison between the apparent color vs. the thickness color-charts obtained for the four different $SiO_2$ thicknesses studied here. If a different $SiO_2$ thickness is used, a new calibration measurement, like that shown in Figure 1, has to be carried out again for the desired $SiO_2$ thickness.

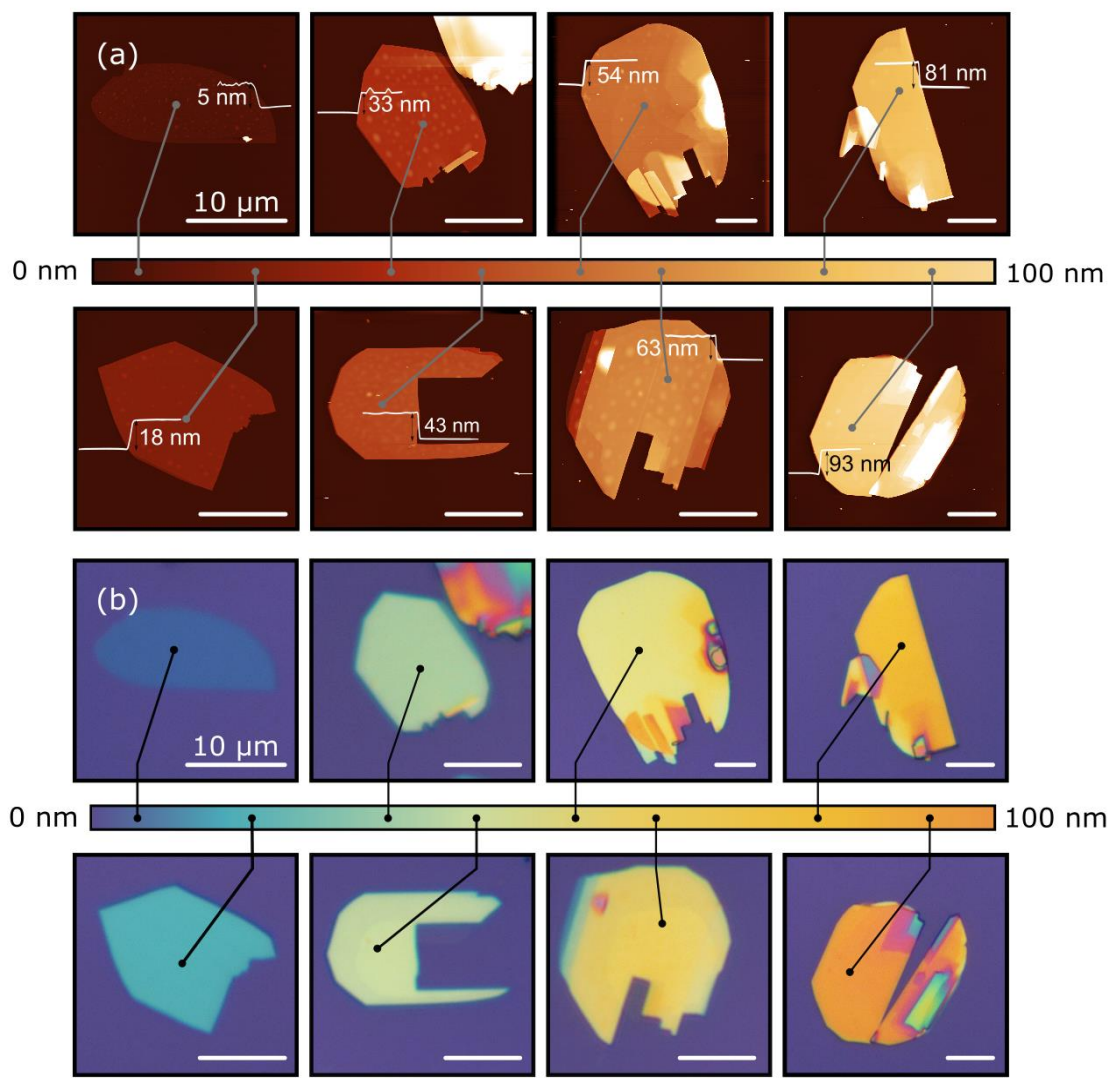

**Figure 1.** Thickness dependent apparent color of MoO₃ flakes. (**a**) AFM measurements of the exfoliated flakes with different thickness placed on a 297 nm SiO₂/Si substrate. (**b**) Optical images of the flakes and a colorbar with the apparent color of flakes with thickness from 5 nm up to 93 nm (from 7 to ~130 layers). Scale bars: 10 μm.

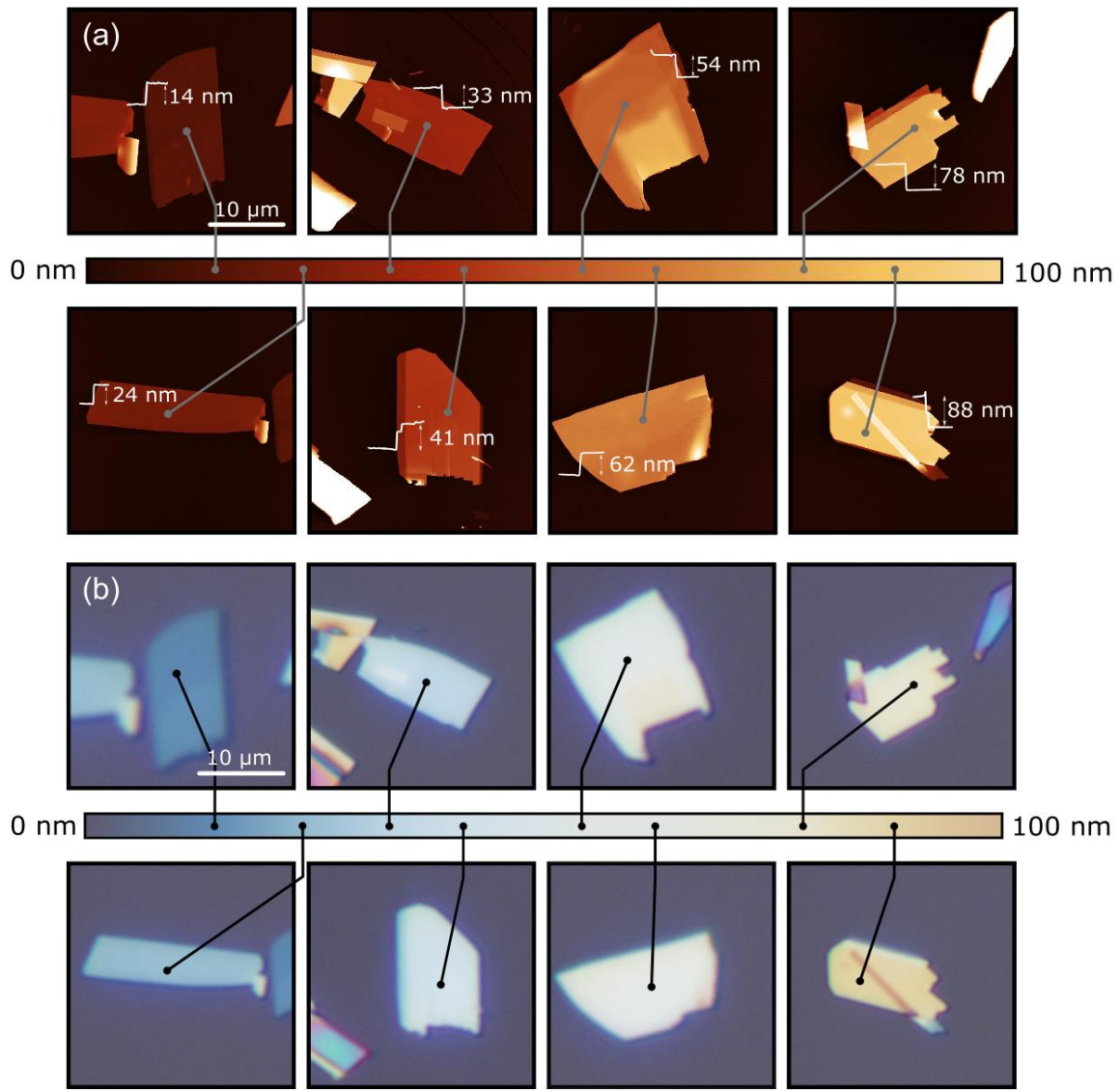

**Figure 2.** Thickness dependent apparent color of MoO₃ flakes. (**a**) AFM measurements of the exfoliated flakes with different thickness placed on an 88 nm SiO₂/Si substrate. (**b**) Optical images of the flakes and a colorbar with the apparent color of flakes with thickness up to 88 nm. Scale bars: 10 μm.

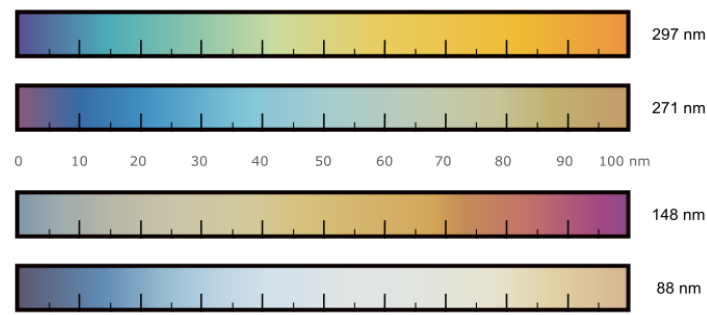

**Figure 3.** Thickness dependent apparent color of MoO₃ flakes on SiO₂/Si substrates with different oxide capping layer thickness.

Further, we quantitatively analyze the reflection spectra of MoO₃ flakes to measure their thickness more accurately, similarly to previous works in transition metal dichalcogenides, muscovite mica and black phosphorus [44,49–51]. Differential reflectance spectra are acquired in normal incidence with a modified metallurgical microscope (BA 310 MET-T, Motic), details in Reference [44]. A spectrum is first acquired onto the bare substrate ($I_s$) and then onto the flake ($I_f$) and the optical contrast ($C$) can be calculated as:

$$C = \frac{I_f - I_s}{I_f + I_s} \tag{1}$$

In Figure 4a, we sketch the optical system, indicating the different optical media, used to model the optical contrast spectra. The refractive index values of the different media used for the model are shown in Figure S3a. Figure 4b shows the experimental optical contrast obtained for different values of thickness of MoO₃.

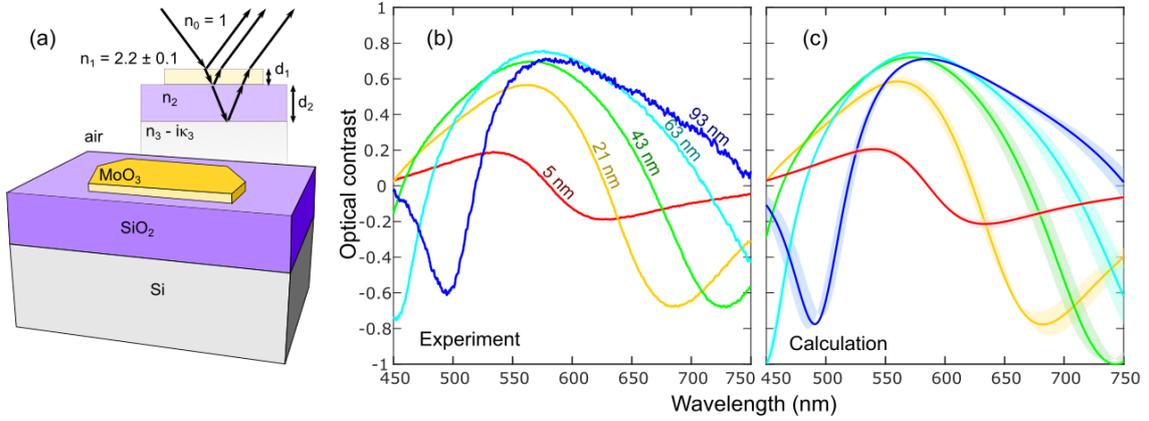

**Figure 4.** (**a**) Optical model used to calculate the MoO₃ optical contrast. (**b**) Optical contrast spectra measured on MoO₃ flakes, on a 297 nm SiO₂/Si substrate, of different thickness. (**c**) Calculated optical contrast (solid lines) using the Fresnel law-based model[52]. The shaded area accounts for an uncertainty of ±1 nm in the thickness of the flake[53].

The optical contrast of this kind of multilayer optical system can be calculated with high accuracy using a Fresnel law-based model [54]. The reflection coefficient in a four media Fresnel model is expressed as [55]:

$$r_4 = \frac{r_{01}e^{i(\phi_1+\phi_2)} + r_{12}e^{-i(\phi_1-\phi_2)} + r_{23}e^{-i(\phi_1+\phi_2)} + r_{01}r_{12}r_{23}e^{i(\phi_1-\phi_2)}}{e^{i(\phi_1+\phi_2)} + r_{01}r_{12}e^{-i(\phi_1-\phi_2)} + r_{01}r_{23}e^{-i(\phi_1+\phi_2)} + r_{12}r_{23}e^{i(\phi_1-\phi_2)}} \tag{2}$$

where sub index 0 refers to air, 1 to MoO₃, 2 to SiO₂ and 3 to Si. Assuming normal incidence, $\Phi_i = 2\pi\tilde{n}_i d_i/\lambda$ is the phase shift induced by the propagation of the light beam in the media $i$, being $\tilde{n}_i, d_i$ and $\lambda$ the complex refractive index, thickness of the media and wavelength, respectively; $r_{ij} = (\tilde{n}_i - \tilde{n}_j)/(\tilde{n}_i + \tilde{n}_j)$ is the Fresnel coefficient at the interface between the media $i$ and $j$.

The reflection coefficient in a three media Fresnel model (i.e., the case of the bare substrate without the MoO₃ flake) is expressed as:

$$r_3 = \frac{r_{01} + r_{12}e^{-i2\phi_1}}{1 + r_{01}r_{12}e^{-i2\phi_1}} \tag{3}$$

where sub index 0 is air, 1 is SiO₂ and 2 is Si. With these equations, one can calculate the optical contrast by firstly calculating the reflected intensity:

$$R_k = |\overline{r_k}r_k| \, , \forall \, k = 3, 4 \tag{4}$$

Then the optical contrast can be defined through the following operation that correlates the reflected intensity by the bare substrate ($R_3$) with the reflected intensity by the MoO₃ flake ($R_4$) as:

$$C = \frac{R_4 - R_3}{R_4 + R_3} \qquad (5)$$

Interestingly, we found that one can accurately reproduce the experimental optical contrast spectra by simply assuming a refractive index of MoO₃ of $\tilde{n}_1 = 2.2 - i0$. Note that, according to Reference [35], the band structure of MoO₃ has a negligible thickness dependence and thus we do not expect the refractive index to depend on the flake thickness. The results of the calculated spectra with this model are depicted in Figure 4c. The real part of the refractive index of bulk MoO₃ $n_{MoO_3}$ spans in the literature from 1.8 to 2.3 [14,56,57]. Since the literature values of the extinction coefficient in the visible range are very low, we neglect that term in our calculations, still providing a good fit to the experimental data. As verification, Figure S3b shows the complex refractive index as a function of wavelength, $\tilde{n} = n + i\kappa$, measured with ellipsometry of polycrystalline MoO₃ films showing values of $n = 2$–2.3 in the visible part of the spectrum (from 1.5 to 3 eV) and a negligible value of $\kappa$ in the same region (from 0.02 to 0.1). This ellipsometry measurement on polycrystalline MoO₃ verifies that the assumed refractive index for single crystalline MoO₃ flakes ($\tilde{n}_1 = 2.2 - i0$) is reasonable.

Some of the features present in the optical contrast spectra, like the local maxima and minima, strongly depend on the thickness of the MoO₃ medium. In fact, the shape of the optical contrast spectra, including the maxima and minima features, arises from the interference colors effect. It is therefore clear that these features will depend on the thickness of the MoO₃ flake as the optical paths of the light beams passing through flakes with different thicknesses will be different. Therefore, one can evaluate the thickness of a MoO₃ flake by calculating the optical contrast according to Equations (2)–(5) for different thicknesses of MoO₃ and determining the best fit by minimum squares.

Figure 5a compares the optical contrast measured for a MoO₃ flake (16.5 nm thick according to the AFM) with the optical contrast calculated assuming a thickness in the range of 0–100 nm. The best match is obtained for a thickness of 13.5 nm. The inset in Figure 5a shows the square of the difference between the measured contrast and the calculated one as a function of the thickness assumed for the calculation. The plot shows a well-defined minimum at a thickness of 13.5 nm.

In order to benchmark this thickness determination method, Figure 5b compares the thickness values measured with AFM for 23 flakes from 5 nm to 100 nm thick with the values obtained following the optical contrast fit method discussed above. In the plot, we include a straight line with a slope of 1 that indicates the perfect agreement between the thickness of MoO₃ nanosheets measured by AFM and the fit to the Fresnel law-based model. The low dispersion along the slope = 1 line indicates the good agreement between the thickness evaluated by both methods. In effect, the calculated linear regression of the data points in Figure 5b has a slope of 1.02.

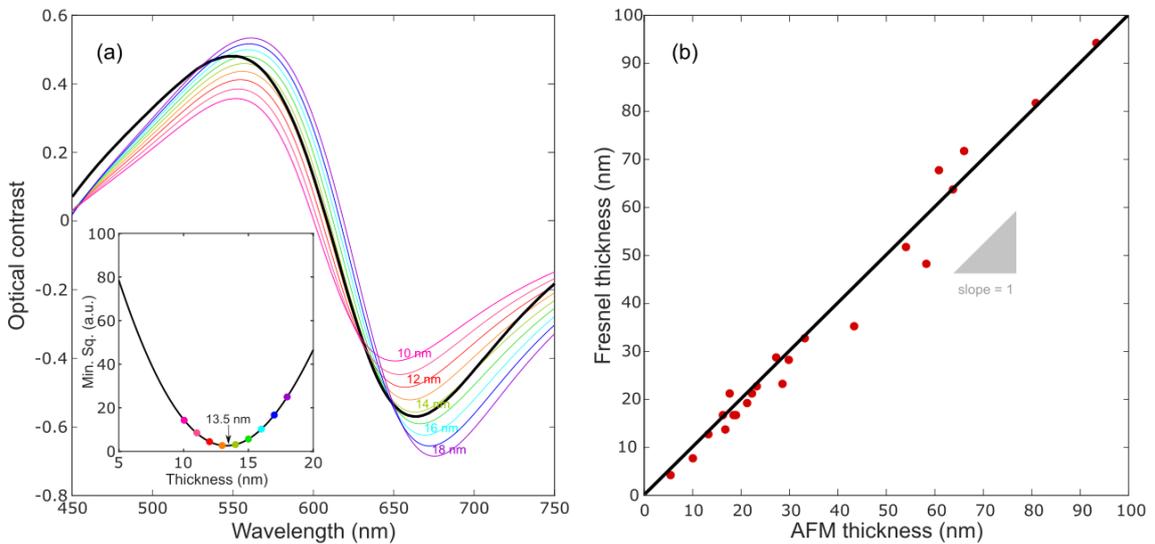

**Figure 5. (a)** Experimental optical contrast (black) and comparison with the calculated ones (color lines) of $MoO_3$ flakes, with thicknesses from 10 to 18 nm, placed onto a substrate of 297 nm of $SiO_2$. Inset of figure: minimum square value for different values of thickness of the flakes. **(b)** Comparison of thickness measured with AFM and the Fresnel model. Experimental data represented as red dots and line in black has a slope of 1.

## 4. Conclusions

In summary, we provided two fast and non-destructive complementary methods to evaluate the number of layers of $MoO_3$ nanosheets using optical microscopy. First, one can get a coarse estimation of the thickness (with ±3 nm of uncertainty) by comparing the apparent color of the flakes with a pre-calibrated color-chart. This method is very fast, but it requires the previous calibration of a color-chart that depends on the $SiO_2$ capping layer thickness (nonetheless, we provide pre-calibrated color-charts for four commonly used $SiO_2$ capping layer thicknesses). The second method is based on the measurement of the optical contrast spectrum of the $MoO_3$ flake under study and the subsequent fit to a Fresnel law-based model that includes the optical constants of each medium. This method requires a modification of the optical microscope to allow for differential reflectance measurements, but it can provide lower uncertainty (±2 nm of median standard deviation for the samples on 297 nm of $SiO_2$), and it can be easily employed for flakes transferred on $SiO_2$/Si substrates with different $SiO_2$ thicknesses. We believe that the development of these thickness determination methods can be very helpful for the community working on $MoO_3$ as it can effectively speed up the research on this 2D material.

**Author Contributions:** S.P. fabricated the $MoO_3$ flake samples and performed the experimental work. A.M.-J. and R.S.G. performed the ellipsometry measurements on the polycrystalline $MoO_3$ film. C.M. and A.C.-G. supervised and designed the experiments. S.P. and A.C.-G. drafted the firsts version of the manuscript. All authors contributed to writing the final version of the manuscript.

**Funding:** This project has received funding from the European Research Council (ERC) under the European Union's Horizon 2020 research and innovation programme (grant agreement n° 755655, ERC-StG 2017 project 2D-TOPSENSE), the European Commission, under the Graphene Flagship (Core 3, grant number 881603), the Spanish Ministry of Economy, Industry and Competitiveness through the grant MAT2017-87134-C2-2-R and partially funded by the Spanish Ministerio de Ciencia e Innovación through grant RTI2018-096498-B-I00 (MCIU/AEI/FEDER, UE). SP acknowledges the fellowship PRE2018-084818.

## References

1.    Novoselov, K.S.; Geim, A.K.; Morozov, S. V; Jiang, D.; Zhang, Y.; Dubonos, S. V; Grigorieva, I. V; Firsov, A.A. Electric field effect in atomically thin carbon films. *Science (80-. ).* **2004**, *306*, 666–669, doi:10.1126/science.1102896.

2.    Burzurí, E.; Prins, F.; van der Zant, H.S.J. Characterization of Nanometer-Spaced Few-Layer Graphene Electrodes. *Graphene* **2012**, *01*, 26–29, doi:10.4236/graphene.2012.12004.

3.    Tsai, D.S.; Liu, K.K.; Lien, D.H.; Tsai, M.L.; Kang, C.F.; Lin, C.A.; Li, L.J.; He, J.H. Few-layer MoS2 with

high broadband photogain and fast optical switching for use in harsh environments. *ACS Nano* **2013**, *7*, 3905–3911, doi:10.1021/nn305301b.

4.  Dean, C.R.; Young, A.F.; Meric, I.; Lee, C.; Wang, L.; Sorgenfrei, S.; Watanabe, K.; Taniguchi, T.; Kim, P.; Shepard, K.L.; et al. Boron nitride substrates for high-quality graphene electronics. **2010**, *5*, 0–4, doi:10.1038/nnano.2010.172.

5.  Akhtar, M.; Anderson, G.; Zhao, R.; Alruqi, A.; Mroczkowska, J.E.; Sumanasekera, G.; Jasinski, J.B. Recent advances in synthesis , properties , and applications of phosphorene. *npj 2D Mater. Appl.* **2017**, 1–12, doi:10.1038/s41699-017-0007-5.

6.  de Castro, I.A.; Datta, R.S.; Ou, J.Z.; Castellanos-Gomez, A.; Sriram, S.; Daeneke, T.; Kalantar-zadeh, K. Molybdenum Oxides – From Fundamentals to Functionality. *Adv. Mater.* **2017**, *29*, 1–31, doi:10.1002/adma.201701619.

7.  Stanford, M.G.; Rack, P.D.; Jariwala, D. Emerging nanofabrication and quantum con fi nement techniques for 2D materials beyond graphene. *npj 2D Mater. Appl.* **2018**, doi:10.1038/s41699-018-0065-3.

8.  Naguib, M.; Mochalin, V.N.; Barsoum, M.W.; Gogotsi, Y. 25th Anniversary Article : MXenes : A New Family of Two-Dimensional Materials. **2013**, 992–1005, doi:10.1002/adma.201304138.

9.  Duan, X.; Wang, C.; Pan, A.; Yu, R.; Duan, X. Two-dimensional transition metal dichalcogenides as atomically thin semiconductors : opportunities and challenges. *Chem. Soc. Rev.* **2015**, doi:10.1039/C5CS00507H.

10. Yu, X.; Marks, T.J.; Facchetti, A. Metal oxides for optoelectronic applications. *Nat. Mater.* **2016**, *15*, 383–396, doi:10.1038/nmat4599.

11. Cai, L.; McClellan, C.J.; Koh, A.L.; Li, H.; Yalon, E.; Pop, E.; Zheng, X. Rapid flame synthesis of atomically thin MoO3 down to monolayer thickness for effective hole doping of WSe2. *Nano Lett.* **2017**, *17*, 3854–3861, doi:10.1021/acs.nanolett.7b01322.

12. Kim, J.H.; Dash, J.K.; Kwon, J.; Hyun, C.; Kim, H.; Ji, E.; Lee, G.-H. van der Waals epitaxial growth of single crystal $\alpha$ -MoO 3 layers on layered materials growth templates. *2D Mater.* **2018**, *6*, 015016, doi:10.1088/2053-1583/aaedc8.

13. Balendhran, S.; Walia, S.; Nili, H.; Ou, J.Z.; Zhuiykov, S.; Kaner, R.B.; Sriram, S.; Bhaskaran, M.; Kalantar-Zadeh, K. Two-dimensional molybdenum trioxide and dichalcogenides. *Adv. Funct. Mater.* **2013**, *23*, 3952–3970, doi:10.1002/adfm.201300125.

14. Vos, M.F.J.; Macco, B.; Thissen, N.F.W.; Bol, A.A.; Kessels, W.M.M. (Erwin) Atomic layer deposition of molybdenum oxide from (N t Bu)2 (NMe2)2Mo and O2 plasma. *J. Vac. Sci. Technol. A Vacuum, Surfaces, Film.* **2016**, *34*, 01A103, doi:10.1116/1.4930161.

15. Hussain, O.M.; Rao, K.S. Characterization of activated reactive evaporated MoO3 thin films for gas sensor applications. *Mater. Chem. Phys.* **2003**, *80*, 638–646, doi:10.1016/S0254-0584(03)00101-9.

16. Zhang, W.B.; Qu, Q.; Lai, K. High-mobility transport anisotropy in few-layer MoO3 and its origin. *ACS Appl. Mater. Interfaces* **2017**, *9*, 1702–1709, doi:10.1021/acsami.6b14255.

17. You, H.; Dai, Y.; Zhang, Z.; Ma, D. Improved performances of organic light-emitting diodes with metal oxide as anode buffer. *J. Appl. Phys.* **2007**, *101*, 1–4, doi:10.1063/1.2430511.

18. Shin, W.J.; Lee, J.Y.; Kim, J.C.; Yoon, T.H.; Kim, T.S.; Song, O.K. Bulk and interface properties of molybdenum trioxide-doped hole transporting layer in organic light-emitting diodes. *Org. Electron.* **2008**, *9*, 333–338, doi:10.1016/j.orgel.2007.12.001.

19. Tseng, Y.C.; Mane, A.U.; Elam, J.W.; Darling, S.B. Ultrathin molybdenum oxide anode buffer layer for

organic photovoltaic cells formed using atomic layer deposition. *Sol. Energy Mater. Sol. Cells* **2012**, *99*, 235–239, doi:10.1016/j.solmat.2011.12.004.

20.  Zhao, Y.; Nardes, A.M.; Zhu, K.; Zhao, Y.; Nardes, A.M.; Zhu, K. Effective hole extraction using MoOx-Al contact in perovskite CH3NH3PbI3 solar cells Effective hole extraction using MoOx-Al contact in perovskite CH3NH3PbI3 solar cells. **2014**, *213906*, doi:10.1063/1.4880899.

21.  Battaglia, C.; Nicolás, S.M. De; Wolf, S. De; Yin, X.; Zheng, M.; Ballif, C.; Wolf, S. De; Yin, X. Silicon heterojunction solar cell with passivated hole selective MoOx contact Silicon heterojunction solar cell with passivated hole selective MoOx contact. **2014**, *113902*, doi:10.1063/1.4868880.

22.  Gurlo, A.; Bârsan, N.; Ivanovskaya, M.; Weimar, U.; Göpel, W. In2O3 and MoO3-In2O3 thin film semiconductor sensors: Interaction with NO2 and O3. *Sensors Actuators, B Chem.* **1998**, *47*, 92–99, doi:10.1016/S0925-4005(98)00033-1.

23.  Ferroni, M.; Guidi, V.; Martinelli, G.; Nelli, P.; Sacerdoti, M.; Sberveglieri, G. Characterization of a molybdenum oxide sputtered thin film as a gas sensor. **1997**, *307*, 148–151.

24.  Quevedo-Lopez, M.A.; Reidy, R.F.; Orozco-Teran, R.A.; Mendoza-Gonzalez, O.; Ramirez-Bon, R. Enhancement of the photochromic and thermochromic properties of molybdenum oxide thin films by a cadmium sulfide underlayer. *J. Mater. Sci. Mater. Electron.* **2000**, *11*, 151–155, doi:10.1023/A:1008933632515.

25.  Scarminio, J.; Lourenço, A.; Gorenstein, A. Electrochromism and photochromism in amorphous molybdenum oxide films. *Thin Solid Films* **1997**, *302*, 66–70, doi:10.1016/S0040-6090(96)09539-9.

26.  Ivanova, T.; Gesheva, K.A.; Popkirov, G.; Ganchev, M.; Tzvetkova, E. Electrochromic properties of atmospheric CVD MoO3 and MoO3-WO3 films and their application in electrochromic devices. **2005**, *119*, 232–239, doi:10.1016/j.mseb.2004.12.084.

27.  Vao, J.N.; Hashimoto, K.; Fujishima, A. *Photochromism induced in an electrolytically pretreated MoO3 thin film by visible light*; 1991; Vol. 254;.

28.  He, T.; Yao, J. Photochromism of molybdenum oxide. **2003**, *4*, 125–143, doi:10.1016/S1389-5567(03)00025-X.

29.  Thuy, T.; Pham, P.; Hoang, P.; Nguyen, D.; Vo, T.T.; Luu, C.L. Preparation of NO-doped b-MoO3 and its methanol oxidation property. *Mater. Chem. Phys.* **2016**, 1–7, doi:10.1016/j.matchemphys.2016.09.048.

30.  Mizushima, T.; Moriya, Y.; Huu, N.; Phuc, H.; Ohkita, H.; Kakuta, N. Soft chemical transformation of α-MoO3 to β-MoO3 as a catalyst for vapor-phase oxidation of methanol. **2011**, *13*, 10–13, doi:10.1016/j.catcom.2011.06.012.

31.  Mizushima, T.; Fukushima, K.; Ohkita, H.; Kakuta, N. Synthesis of b-MoO3 through evaporation of HNO3-added molybdic acid solution and its catalytic performance in partial oxidation of methanol. **2007**, *326*, 106–112, doi:10.1016/j.apcata.2007.04.006.

32.  Phuong, P.T.T.; Duy, N.P.H. Facile synthesis of a green metastable MoO3 for the selective oxidation of methanol to formaldehyde. *React. Kinet. Mech. Catal.* **2016**, *117*, 161–171, doi:10.1007/s11144-015-0938-9.

33.  Machiels, C.J.; Cheng, W.H.; Chowdhry, U.; Farneth, W.E.; Hong, F.; Sleight, A.W.; Station, E. The effect of the structure of molybdenum oxides on the selective oxidation of methanol. **1986**, *25*, 249–256.

34.  Zhang, J.; Pan, Y.; Chen, Y.; Lu, H. Plasmonic molybdenum trioxide quantum dots with noble metal-comparable surface enhanced Raman scattering. *J. Mater. Chem. C* **2018**, *6*, 2216–2220, doi:10.1039/c7tc04807f.

35.  Molina-Mendoza, A.J.; Lado, J.L.; Island, J.O.; Niño, M.A.; Aballe, L.; Foerster, M.; Bruno, F.Y.; López-

Moreno, A.; Vaquero-Garzon, L.; Van Der Zant, H.S.J.; et al. Centimeter-Scale Synthesis of Ultrathin Layered MoO3 by van der Waals Epitaxy. *Chem. Mater.* **2016**, *28*, 4042–4051, doi:10.1021/acs.chemmater.6b01505.

36. Kalantar-Zadeh, K.; Tang, J.; Wang, M.; Wang, K.L.; Shailos, A.; Galatsis, K.; Kojima, R.; Strong, V.; Lech, A.; Wlodarski, W.; et al. Synthesis of nanometre-thick MoO3 sheets. *Nanoscale* **2010**, *2*, 429–433, doi:10.1039/b9nr00320g.

37. Zheng, B.; Wang, Z.; Chen, Y.; Zhang, W.; Li, X. Centimeter-sized 2D $\alpha$-MoO3 single crystal: Growth, Raman anisotropy, and optoelectronic properties. *2D Mater.* **2018**, *5*, doi:10.1088/2053-1583/aad2ba.

38. Reed, B.W.; Williams, D.R.; Moser, B.P.; Koski, K.J. Chemically Tuning Quantized Acoustic Phonons in 2D Layered MoO3 Nanoribbons. *Nano Lett.* **2019**, *19*, 4406–4412, doi:10.1021/acs.nanolett.9b01068.

39. Zheng, Q.; Ren, P.; Peng, Y.; Zhou, W.; Yin, Y.; Wu, H.; Gong, W.; Wang, W.; Tang, D.; Zou, B. In-Plane Anisotropic Raman Response and Electrical Conductivity with Robust Electron-Photon and Electron-Phonon Interactions of Air Stable MoO2 Nanosheets. *J. Phys. Chem. Lett.* **2019**, *10*, 2182–2190, doi:10.1021/acs.jpclett.9b00455.

40. Yuan, H.; Liu, X.; Afshinmanesh, F.; Li, W.; Xu, G.; Sun, J.; Lian, B.; Curto, A.G.; Ye, G.; Hikita, Y.; et al. Polarization-sensitive broadband photodetector using a black phosphorus vertical p-n junction. *Nat. Nanotechnol.* **2015**, *10*, 707–713, doi:10.1038/nnano.2015.112.

41. Wang, X.; Li, Y.; Huang, L.; Jiang, X.W.; Jiang, L.; Dong, H.; Wei, Z.; Li, J.; Hu, W. Short-Wave Near-Infrared Linear Dichroism of Two-Dimensional Germanium Selenide. *J. Am. Chem. Soc.* **2017**, *139*, 14976–14982, doi:10.1021/jacs.7b06314.

42. Ma, W.; Alonso-gonzález, P.; Li, S.; Nikitin, A.Y.; Yuan, J.; Martín-sánchez, J.; Sriram, S.; Kalantar-zadeh, K.; Lee, S.; Hillenbrand, R.; et al. In-plane anisotropic and ultra-low-loss polaritons in a natural van der Waals crystal. *Nature* **2018**, doi:10.1038/s41586-018-0618-9.

43. Horcas, I.; Fernández, R.; Gómez-Rodríguez, J.M.; Colchero, J.; Gómez-Herrero, J.; Baro, A.M. WSXM: A software for scanning probe microscopy and a tool for nanotechnology. *Rev. Sci. Instrum.* **2007**, *78*, doi:10.1063/1.2432410.

44. Frisenda, R.; Niu, Y.; Gant, P.; Molina-Mendoza, A.J.; Schmidt, R.; Bratschitsch, R.; Liu, J.; Fu, L.; Dumcenco, D.; Kis, A.; et al. Micro-reflectance and transmittance spectroscopy: A versatile and powerful tool to characterize 2D materials. *J. Phys. D. Appl. Phys.* **2017**, *50*, doi:10.1088/1361-6463/aa5256.

45. Li, X.; Cai, W.; An, J.; Kim, S.; Nah, J.; Yang, D.; Piner, R.; Velamakanni, A.; Jung, I.; Tutuc, E.; et al. Large-area synthesis of high-quality and uniform graphene films on copper foils. *Science* **2009**, *324*, 1312–4, doi:10.1126/science.1171245.

46. Li, H.; Wu, J.; Huang, X.; Lu, G.; Yang, J.; Lu, X.; Xiong, Q.; Zhang, H. Rapid and reliable thickness identification of two-dimensional nanosheets using optical microscopy. *ACS Nano* **2013**, *7*, 10344–53, doi:10.1021/nn4047474.

47. Papadopoulos, N.; Frisenda, R.; Biele, R.; Flores, E.; Ares, J.R.; Sánchez, C.; Van Der Zant, H.S.J.; Ferrer, I.J.; D'Agosta, R.; Castellanos-Gomez, A. Large birefringence and linear dichroism in TiS3 nanosheets. *Nanoscale* **2018**, *10*, doi:10.1039/c8nr03616k.

48. Gant, P.; Ghasemi, F.; Maeso, D.; Munuera, C.; López-Elvira, E.; Frisenda, R.; De Lara, D.P.; Rubio-Bollinger, G.; Garcia-Hernandez, M.; Castellanos-Gomez, A. Optical contrast and refractive index of natural van der Waals heterostructure nanosheets of franckeite. *Beilstein J. Nanotechnol.* **2017**, *8*, doi:10.3762/bjnano.8.235.

49.   Rubio-Bollinger, G.; Guerrero, R.; de Lara, D.; Quereda, J.; Vaquero-Garzon, L.; Agraït, N.; Bratschitsch, R.; Castellanos-Gomez, A. Enhanced Visibility of $MoS_2$, $MoSe_2$, $WSe_2$ and Black-Phosphorus: Making Optical Identification of 2D Semiconductors Easier. *Electronics* **2015**, *4*, 847–856, doi:10.3390/electronics4040847.

50.   Niu, Y.; Gonzalez-Abad, S.; Frisenda, R.; Marauhn, P.; Drüppel, M.; Gant, P.; Schmidt, R.; Taghavi, N.; Barcons, D.; Molina-Mendoza, A.; et al. Thickness-Dependent Differential Reflectance Spectra of Monolayer and Few-Layer MoS2, MoSe2, WS2 and WSe2. *Nanomaterials* **2018**, *8*, 725, doi:10.3390/nano8090725.

51.   Castellanos-Gomez, A.; Wojtaszek, M.; Tombros, N.; Agrait, N.; van Wees, B.J.; Rubio-Bollinger, G. Atomically thin mica flakes and their application as ultrathin insulating substrates for graphene. *Small* **2011**, *7*, 2491–2497, doi:10.1002/smll.201100733.

52.   Ares, P.; Zamora, F.; Gomez-Herrero, J. Optical Identification of Few-Layer Antimonene Crystals. *ACS Photonics* **2017**, *4*, 600–605, doi:10.1021/acsphotonics.6b00941.

53.   C., R. 54. B.; A.M., A.; C., A.; A., A.-L.; R., A.; J., A.; N., B.; L., B.; J., B.; Bartali Production and processing of graphene and related materials. *2D Mater.* **2020**, *7*, 22001.

54.   E, Hecht; A, Z. *Optics*; Hecht, E., Ed.; 4th ed.; Addison Wesley Longman: Amsterdam, 2002; ISBN 0-321-18878-0.

55.   Anders, H. Thin films in optics. *Optom. Vis. Sci.* **1968**, *45*, 781, doi:10.1097/00006324-196811000-00013.

56.   Szekeres, A.; Ivanova, T.; Gesheva, K. Spectroscopic ellipsometry study of CVD molybdenum oxide films : effect of temperature. **2002**, 17–20, doi:10.1007/s10008-002-0285-4.

57.   Deb, S.K.; A, P.R.S.L. Physical properties of a transition metal oxide: optical and photoelectric properties of single crystal and thin film molybdenum trioxide. *Proc. R. Soc. London. Ser. A. Math. Phys. Sci.* **1968**, *304*, 211–231, doi:10.1098/rspa.1968.0082.